\documentclass[letterpaper,twocolumn,10pt]{article}
\usepackage{usenix2019_v3}

\usepackage{tikz}
\usepackage{amsmath}

\usepackage{soul}
\usepackage{xspace}
\usepackage{multirow}
\usepackage{algorithm}
\usepackage{algpseudocode} 

\usepackage{float}
\usepackage[labelfont=normalfont,textfont=normalfont]{subcaption}
\usepackage{tablefootnote}
\usepackage{footmisc}

\usepackage{comment}

\usepackage{amssymb}
\usepackage{amsthm}

\newcommand{\defmacro}[2]{%
  \expandafter\def\csname#1\endcsname##{#2}
  \expandafter\soulregister\csname#1\endcsname7
}
\defmacro{ralt}{RALT}
\defmacro{raltfull}{Recent Access Lookup Table}
\defmacro{hotrap}{HotRAP}
\defmacro{hotrapfull}{Hot record Retention And Promotion}
\defmacro{fastdisk}{FD}
\defmacro{fastdiskfull}{Fast Disk}
\defmacro{slowdisk}{SD}
\defmacro{slowdiskfull}{Slow Disk}

\newcommand{\tablefontsize}{\small}

\newcommand{\circled}[1]{\textcircled{\footnotesize{#1}}}

\usepackage[normalem]{ulem}

\defmacro{SpeedupRO}{5.4}
\defmacro{SpeedupRW}{3.8}
\defmacro{MaxSpeedupRORW}{5.4}

\defmacro{OverheadUniformRocksdbTiered}{0.4\%}
\defmacro{OverheadUniformRocksdbTieredCeil}{1\%}

\defmacro{MaxSpeedupTwitterRocksdbTiered}{5.27}

\defmacro{MaxSpeedupTwitter}{1.9}

\defmacro{RaltCpuPortionMin}{3.7\%}
\defmacro{RaltCpuPortionMax}{13.3\%}

\defmacro{RaltIoPortionMin}{5.5\%}
\defmacro{RaltIoPortionMax}{12.7\%}

\defmacro{PromoteAccessedPromotedMore}{173.0$\times$}
\defmacro{PromoteAccessedCompactionMore}{248.4$\times$}

\begin{document}

\date{}

\title{\Large \bf \hotrap{}: Hot Record Retention and Promotion for LSM-trees with Tiered Storage}

\author{
{\rm Jiansheng Qiu}\\
Tsinghua University
\and
{\rm Fangzhou Yuan}\\
Tsinghua University
\and
{\rm Mingyu Gao}\\
Tsinghua University
\and
{\rm Huanchen Zhang}\\
Tsinghua University
}

\maketitle

\begin{abstract}

The multi-level design of Log-Structured Merge-trees (LSM-trees) naturally fits the tiered storage architecture: the upper levels (recently inserted/updated records) are kept in fast storage to guarantee performance while the lower levels (the majority of records) are placed in slower but cheaper storage to reduce cost. However, frequently accessed records may have been compacted and reside in slow storage. Existing algorithms are inefficient in promoting these ``hot'' records to fast storage, leading to compromised read performance. We present \hotrap{}, a key-value store based on RocksDB that can timely promote hot records individually from slow to fast storage and keep them in fast storage while they are hot. \hotrap{} uses an on-disk data structure (a specially-made LSM-tree) to track the hotness of keys and includes three pathways to ensure that hot records reach fast storage with short delays. Our experiments show that \hotrap{} outperforms state-of-the-art LSM-trees on tiered storage by up to \MaxSpeedupRORW{}$\times$ compared to the second best under read-only and read-write-balanced YCSB workloads with common access skew patterns, and by up to \MaxSpeedupTwitter{}$\times$ compared to the second best under Twitter production workloads.

\end{abstract}

\section{Introduction}

Log-Structured Merge-trees (LSM-trees)~\cite{LSM-tree, LSM-tree2} are widely adopted to build key-value stores~\cite{LevelDB, RocksDBGithub, dong2017optimizing, TiKV} and database storage engines~\cite{ScyllaDB, CockroachDB, Cassandra, huang2020tidb, matsunobu2020myrocks} because of their superior write performance. To achieve better cost efficiency, systems tend to leverage the tiered storage by locating the upper levels of the LSM-tree in fast local solid-state drives (SSDs) while storing the lower levels (i.e., the majority of the records) in slower but cheaper cloud storage or hard disk drives (HDDs). Such a storage-tier separation is inherently efficient for the write operations (i.e., inserts, updates, deletes) because the append-only nature of the LSM-tree keeps the most recent writes automatically in the fast storage. 
However, although in typical workloads the records that are frequently accessed are often correlated with those that are frequently updated, the ``read-hot'' set and the ``write-hot'' set may not always fully overlap.
This leads to a majority of the read-hot records sitting in slow storage with higher latency, thus compromising the read performance of the LSM-tree.

The problem can be mitigated by caching frequently accessed records in memory, and prior studies have proposed numerous caching algorithms~\cite{wu2016zexpander, wu2020ac, yang2020leaper, zhang2022halsm, zhou2023calcspar}. However, memory is often a limiting resource for systems~\cite{zhang2016reducing, zhang2018surf, mcallister2021kangaroo}, and the size of hot records can be far larger than the memory capacity (limitation 1). RocksDB, therefore, introduces the secondary cache on fast SSDs for caching recently-accessed data blocks~\cite{RocksDBSecondaryCache, SAS-Cache}. Solutions such as Mutant~\cite{yoonMutantBalancingStorage2018}, LogStore~\cite{menon2020logstore}, and MirrorKV~\cite{wang2023mirrorkv} propose to adjust the placement of blocks/SSTables across storage tiers periodically according to their access frequencies. These approaches, nonetheless, cannot fully leverage the capacity of the expensive fast storage because they move data at a coarse granularity, where many cold records in the identified hot blocks/SSTables are altogether piggybacked to the fast storage (limitation 2). 
Moreover, the above solutions can only promote hot data to fast storage through LSM-tree compactions. To deal with read-heavy workloads where compactions happen infrequently, several systems~\cite{menon2020logstore,PrismDB} allow triggering compactions proactively, but they must wait for the hot records to accumulate in the slow storage before promoting them. Such a delay harms read performance because it could overstep the time window when a record is hot (limitation 3).

\begin{figure}
    \centering
    \includegraphics{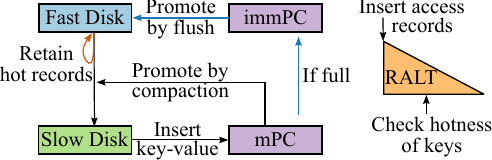}
    \caption{The high-level picture of \hotrap{}. \textnormal{\ralt{} is a small LSM-tree in FD that tracks the hotness of keys. \emph{mPC} and \emph{immPC} stand for the mutable and immutable promotion caches. 
    Hot records are retained in FD during compactions. Records accessed in SD are inserted into the promotion cache and then promoted by compaction or flush if they are hot.}}\label{fig:big-picture}
\end{figure}

In this paper, we present \hotrap{} (\hotrapfull{}), an LSM-tree design based on RocksDB on tiered storage. \hotrap{} can promote hot records timely from the slow disk (abbr. as \emph{SD} hereafter) to the fast disk (abbr. as \emph{FD} hereafter), and retain them in FD as long as they stay hot. \hotrap{} addresses the aforementioned three limitations of previous solutions. 
Figure~\ref{fig:big-picture} shows the big picture of \hotrap{}.

First, instead of tracking the record hotness in memory, \hotrap{} logs each record access in a small specially-made LSM-tree, called \emph{\ralt{}} (\raltfull{}), located in FD (addressing limitation 1). 
\ralt{} tracks the access history for each logged key and maintains a hotness score for the key using exponential smoothing. \ralt{} then evicts low-score keys periodically from itself to stay under a size limit that can be automatically tuned according to the workload.
We choose LSM-tree to implement \ralt{} in order to benefit from its low write latency on disks because inserting access records to it is on the critical path of lookups.

By utilizing \ralt{}, \hotrap{} also supports hot record retention and promotion at the \emph{record level} rather than at the block/SSTable level, thus preventing cold records from being piggybacked to FD (addressing limitation 2).
Also, besides waiting for LSM-tree compactions to retain/promote hot records, \hotrap{} introduces a small in-memory \emph{promotion cache} to cache records read from \slowdisk{} and timely promotes the hot ones (via checking \ralt{}) by flushing them to the top of the data LSM-tree (addressing limitation 3).

More specifically, \hotrap{} provides the following three pathways for hot records to reside in FD: retention, promotion by compaction, and promotion by flush.
First,
during compactions from \fastdisk{} to \slowdisk{},
\hotrap{} checks the hotness of each record in the selected SSTable in \fastdisk{}, and hot records are written back to \fastdisk{} instead of merged down to \slowdisk{}, i.e., \emph{retention}. 
Second, an in-memory mutable promotion cache in \hotrap{} is used to cache accessed records in \slowdisk{}. When the above compaction from \fastdisk{} to \slowdisk{} happens, \hotrap{} also checks the hotness of the records in the mutable promotion cache that are within the compaction key range, and similarly handle these records, writing the hot ones to \fastdisk{} and compacting the others to \slowdisk{}. We call this path \emph{promotion by compaction}. 
Both the hotness checks in these two paths are facilitated by \ralt{} and can be efficiently performed mainly with sequential scans of corresponding key ranges.
Third, when the mutable promotion cache grows too big due to insufficient compactions, we turn it into an immutable promotion cache and trigger \emph{promotion by flush} to bulk-insert the hot records in the immutable promotion cache (also identified via consulting \ralt{}) to $L_0$ of the LSM-tree. Thus we effectively restrict the size of the promotion cache. 
To prevent promotion by flush from overwriting records with a newer version (i.e., promoting a stale record to $L_0$ that has a higher level than the newly updated version), we perform extra checks and carry out a concurrency control mechanism to guarantee correctness.

We evaluated \hotrap{} using YCSB-based workloads and Twitter production traces~\cite{twittertrace} on AWS instances with fast local NVMe SSDs and slower cloud storage. We compare \hotrap{} to the state-of-the-art LSM-trees on tiered storage, e.g., Mutant~\cite{yoonMutantBalancingStorage2018} (promoting at the SSTable level), SAS-Cache~\cite{SAS-Cache} (based on RocksDB with secondary block cache~\cite{RocksDBSecondaryCache}), and PrismDB~\cite{PrismDB} (proactively triggering compactions). \hotrap{} achieves \SpeedupRO{}$\times$ speedup over the second best for read-only YCSB workloads, \SpeedupRW{}$\times$ for read-write-balanced YCSB workloads, and \MaxSpeedupTwitter{}$\times$ for Twitter production workloads, while maintaining competitive performance for write-heavy and update-heavy YCSB workloads. Our experiments also show that \hotrap{} adds $<$ \OverheadUniformRocksdbTieredCeil{} overhead to the plain RocksDB under uniform workloads and is robust against hotspot shifts,
where \ralt{} quickly evicts stale access records and adds keys in the new hotspot into the hot set.

We make three primary contributions in this paper. First, we propose an on-disk data structure (i.e., \ralt{}, a specially-made LSM-tree) for tracking the hotness of key-value records. Second, we design three pathways in a tiered LSM-tree for timely promoting/retaining hot records to/in FD. Our algorithm operates at the record level so that the system can efficiently utilize the limited space of FD. Finally, we implement \hotrap{}, a key-value store based on RocksDB that outperforms state-of-the-art LSM-trees on tiered storage because of its efficient hot record tracking and movement features.

\section{Background \& Related Work}

\subsection{LSM-trees and tiered storage}\label{subsection:lsm-tree}

A Log-Structured Merge-tree (LSM-tree) consists of an in-memory buffer (i.e., MemTable) and multiple levels $L_0,\ldots,L_n$ on disk. The capacity of level $L_i$ is made $T$ times larger than $L_{i-1}$, where $T$ is called the size ratio of the LSM-tree. Records are first inserted into MemTable. When MemTable is full, it becomes immutable and then flushed to $L_0$ as an SSTable (i.e., a file format called Sorted String Table). When level $L_{i-1}$ reaches its capacity, it will trigger the compaction process to merge its content into the next level $L_i$. There are typically two kinds of compaction policies: leveling and tiering. Leveling only allows one sorted run per level while tiering allows multiple. This paper focuses on the leveling policy because it is RocksDB's default choice~\cite{RocksDBGithub}. RocksDB also adopts partial compaction: each compaction picks an SSTables from $L_{i-1}$ whose key range overlaps with a minimal number of SSTables in $L_i$ to merge to the next level. Such a compaction strategy leads to a write amplification of $\approx \frac{nT}{2}$~\cite{dayan2022spooky}.

A lookup first checks the MemTable and then searches the levels from top to bottom until a matching key is found. A block index in memory is used to determine which SSTable data block to search for a particular key in a sorted run. Per-SSTable Bloom filters are used to reduce the number of candidate SSTables to further save I/Os. RocksDB provides snapshot isolation via multiversion concurrency control (MVCC) to prevent compactions from blocking normal read operations. A snapshot in RocksDB, called a superversion, is created after a flush or compaction completes. An old snapshot is garbage collected when no active queries refer to it.

Although many LSM-trees such as RocksDB are initially designed for local SSDs, the multi-level nature of LSM-trees is a good fit for the tiered storage architecture. The upper levels contain the recently inserted and updated records and are therefore kept in the fast storage such as local SSDs because they are more likely to be accessed in the near future. To improve the system's cost efficiency, the majority of records in the lower levels are placed in HDDs~\cite{dong2021evolution, dong2021rocksdb} or low-tier cloud storage based on HDDs~\cite{dong2023disaggregating}. HDDs exhibit higher latency than SSDs, but they are much cheaper. For example, the unit price for a 20TB Seagate Enterprise HDD Exos X20 today is $6.75\times$ cheaper than a 7.68TB SAMSUNG Enterprise SSD PM9A3~\cite{Seagate-Exos-X20,SAMSUNG-PM9A3}. That means a tiered storage with a size ratio of 1:10 based on these hardware can reduce the storage cost by $77\%$ compared to pure SSDs of the same capacity.

\subsection{Hot/cold separation in LSM-trees}

Tracking the access history of (potentially) hot records in memory can incur a large footprint.
According to the Twitter trace~\cite{twittertrace}, for example, $50\%$ of the workloads have a value size smaller than $5\times$ the key size.
We take the AWS EC2 i4i.2xlarge instance for example, which is equipped with 64GB memory and an 1875GB local AWS Nitro SSD~\cite{AWS-instances}. The configuration reflects the typical memory-disk ratio in the industry. If 1TB of its local SSD is used to store hot records, we need at least $1\text{TB} / (1 + 5) = 166.7$GB memory to track those hot keys, which exceeds the physical memory of the instance.

Mutant~\cite{yoonMutantBalancingStorage2018} tracks the access frequencies of SSTables and adjusts their placement periodically to store the hottest SSTables in the faster storage. LogStore~\cite{menon2020logstore} maintains histograms in memory to track the hotness of SSTables and retains/promotes hot SSTables in/to the faster storage. They both separate hot/cold data in the granularity of SSTables, which is too coarse because there can be considerable cold data in the same SSTable that is considered hot.

RocksDB introduces the on-disk secondary cache for data blocks not hot enough for the in-memory block cache~\cite{RocksDBSecondaryCache}.
SAS-Cache~\cite{SAS-Cache} further proposes several optimizations.
However, the granularity of blocks is still too coarse because small objects are prevalent in large-scale systems~\cite{mcallister2021kangaroo}, and there can be many cold tiny records in a hot data block. An alternative solution is to employ a separate on-disk key-value cache (such as~\cite{berg2020cachelib,mcallister2021kangaroo}) on top of the LSM-tree. But separate caches incur extra disk I/O when updating, because it needs to update the key in both cache and LSM-tree. It is also challenging to ensure consistency between the cache and the LSM-tree.

MirrorKV~\cite{wang2023mirrorkv} splits the LSM-tree into the key and value LSM-tree and caches the hottest key SSTables in the faster storage. Additionally, MirrorKV retains the hottest blocks (e.g., 10\%) during compactions from L1 to L2. Similarly, the granularity of SSTables and blocks are both too coarse.

SA-LSM~\cite{SA-LSM} accurately predicts cold data
with survival analysis and demotes cold records from the faster storage to the slower storage. However, SA-LSM does not support promoting hot records back to the faster storage, and the training cost of the survival model is heavy.

PrismDB~\cite{PrismDB} estimates key popularity with the clock algorithm, and the clock bits are indexed with a hash table. Hot records are retained/promoted in/to the fast disk during compactions. However,
the hash table
can consume considerable memory.
Also, the promotion speed of PrismDB is slow because it only promotes during compactions.

2-Tree~\cite{zhou2023two} maintains two B-trees, one in the memory and the other on the disk. It migrates hot records to the in-memory B-tree and cold records to the on-disk B-tree.
However, it does not support tiered storage and the in-memory B-tree consumes huge memory.

\section{Design \& Implementation}

\subsection{Overview}\label{subsection:overview}

\begin{figure*}
\includegraphics{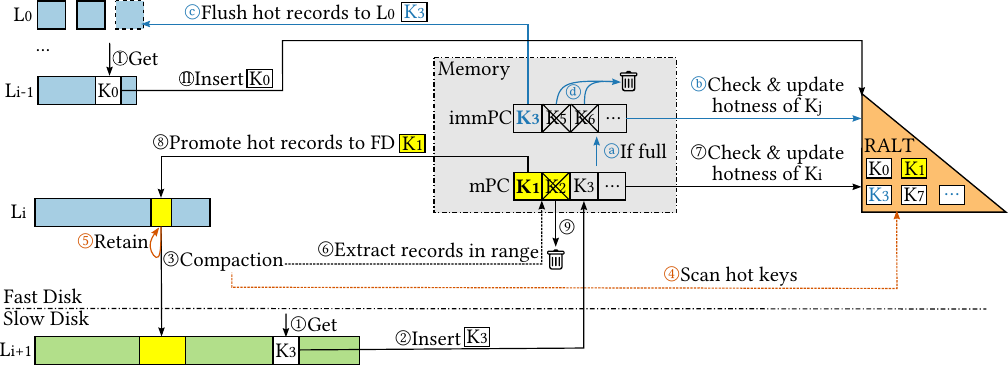}
\centering
\vspace{-0.2cm}
\caption{Overview of \hotrap{}. \textnormal{\emph{mPC} and \emph{immPC} stand for the mutable and immutable promotion caches. Solid arrows are data flow. Dashed arrows are control flow. 
The accessed keys in \slowdisk{} are firstly inserted into the mutable promotion cache (\circled{1} to \circled{2}). 
Hot records are retained in \fastdisk{} during compactions (\circled{3} to \circled{5}). 
A compaction can piggyback hot records in its range to \fastdisk{} (\circled{6} to \circled{9}). 
If the mutable promotion cache becomes full, hot records in it will be flushed to Level 0 (\circled{a} to \circled{e}).}}\label{fig:overview}
\end{figure*}

The overview of \hotrap{} is shown in Figure~\ref{fig:overview}. There are two components to facilitate retention and promotion: \raltfull{} (\emph{\ralt{}}) and the promotion cache. 
\ralt{} is responsible for tracking the hotness of records. It maintains a set of hot keys (without values) that are worth promoting and retaining.
The promotion cache consists of a mutable promotion cache and a list of immutable promotion caches. Records read from \slowdisk{} are inserted into the mutable promotion cache. The mutable promotion cache is turned into immutable when full. Immutable promotion caches are flushed to disk as soon as possible.
The mutable promotion cache resides between the last level of \fastdisk{} and the first level of \slowdisk{}.
To read a key, \hotrap{} first searches in MemTables and levels in \fastdisk{}, then searches in the mutable promotion cache, and at last searches levels in \slowdisk{}.

When a record in \fastdisk{} is accessed (\circled{I}), its key will be inserted into \ralt{} to record the access (\circled{II}). When a record in \slowdisk{} is accessed (\circled{1}), \hotrap{} first inserts the key into the mutable promotion cache (\circled{2}) (unless checks in \S\ref{subsection:check-before-promoting-to-cache} fail),
and the key's hotness information in \ralt{} is updated later.

\textbf{Retention}.
During compactions from \fastdisk{} to \slowdisk{} (\circled{3}),
\hotrap{} checks the hotness of each record to be compacted, and retains the hot ones in \fastdisk{}. It does so by constructing a \ralt{} iterator whose range is the key range of \fastdisk{}'s input SSTables. This iterator produces the hot records identified by \ralt{} in the order of keys. Therefore, we can advance the compaction iterator and the \ralt{} iterator in a sort-merge manner (\circled{4}), to decide the records to retain and write them to new SSTables in \fastdisk{} (\circled{5}). The I/O incurred by the \ralt{} iterator is small because \ralt{} does not store values of \hotrap{} records. 

\textbf{Promotion by compaction}.
During the compaction between \fastdisk{} and \slowdisk{}, we also would like to promote the hot records in \slowdisk{} to \fastdisk{}. 
Recall that the accessed records in \slowdisk{} are cached in the in-memory promotion cache.
We extract the records from the mutable promotion cache that fall within the compaction key range (\circled{6}), so that they could be handled together with this compaction. For the example in Figure~\ref{fig:overview}, $K_1$ and $K_2$ are extracted. 
\hotrap{} consults \ralt{} about whether each key is hot, and updates the hotness score in \ralt{} accordingly (\circled{7}). 
Hot records ($K_1$) are written to the last level in \fastdisk{} (\circled{8}). Cold records ($K_2$) are dropped (\circled{9}) and future lookups to them would go to \slowdisk{}.
Actually, promotion by compaction could also be triggered by a compaction to a level in \fastdisk{}. However, doing promotion during compactions at higher levels of \fastdisk{} would require complex version control, because the records in the promotion cache have older versions.

\textbf{Promotion by flush}. For read-heavy workloads, there may not be enough compactions. In this case, promotion by compaction is insufficient to keep the mutable promotion cache small. Therefore, when the size of the mutable promotion cache grows to the target size of SSTables (64 MiB by default), \hotrap{} converts it to an immutable promotion cache, and a new mutable promotion cache will be created (\circled{a}). For the example in Figure~\ref{fig:overview}, $K_1$ and $K_2$ have been promoted by compaction. Therefore, only $K_3, K_5, \ldots$ are packed into an immutable promotion cache.
The immutable promotion cache consults \ralt{} whether the records are hot (\circled{b}). 
Hot records ($K_3$) will be flushed to $L_0$ in \fastdisk{} (\circled{c}), while cold records ($K_5, K_6$) are dropped (\circled{d}).
To avoid creating tiny SSTables in $L_0$ which will trigger compactions from $L_0$ to $L_1$ prematurely, if the total size of hot records is less than half of the target SSTable size, we insert them back into the mutable promotion cache instead of flushing to $L_0$.

\subsection{\raltfull{} (\ralt{})}\label{subsection:ralt}

\begin{figure}
\centering
\includegraphics{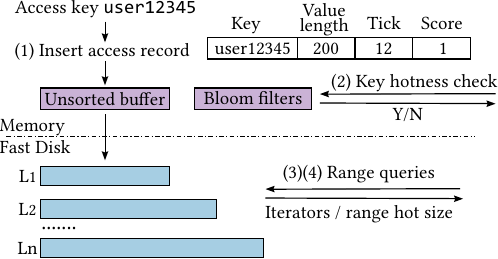}
\vspace{-0.5cm}
\caption{\ralt{} structure. 
\textnormal{Suppose key \texttt{user12345}, whose value length is 200B, is accessed in \hotrap{} and an access record is inserted to \ralt{}. 
The figure shows the \ralt{} access record format, as well as the four supported operations (1) to (4).
The current time slice sequence number for exponential smoothing is 12. 
The \hotrap{} size is len(\texttt{user12345}) + 200 = 209 bytes.
The physical size is $(9 + 4) + (4\times 3) = 25$ bytes, where we use 4 bytes for the length of the key, and 4 bytes each for the value length, tick, and score.}}\label{fig:ralt-overview}
\end{figure}

\ralt{} is a lightweight LSM-tree on \fastdisk{}, logging the accesses to records in \hotrap{}. Its structure is in Figure~\ref{fig:ralt-overview}. 
Each access record in \ralt{} consists of the key, the length of the value (instead of the value itself), and the scoring metadata tick and score. Keys are considered hot if their scores are greater than a threshold. 
We distinguish the size of the original key-value record in \hotrap{} (called the \emph{\hotrap{} size}) and the size of the \ralt{} access record itself (called the \emph{physical size}).
The total \hotrap{} size of hot records is called the \emph{hot set size}. 
\ralt{} has two parameters: the \emph{hot set size limit} and the \emph{physical size limit}, which constrain the total size of hot records and the disk usage of \ralt{} itself, respectively.
We set their values based on workload characteristics in \S\ref{subsection:autotune}.

\ralt{} supports four operations in Figure~\ref{fig:ralt-overview}: (1) insert access records; (2) check if a key is hot; (3) calculate the hot set size in a range; (4) scan hot keys in a range.
When \hotrap{} inserts an access record into \ralt{}, it first goes into an in-memory unsorted buffer. We use an unsorted buffer to improve performance because a sorted memory table does not benefit much: if a key is accessed again while its last access record is in the buffer, it should be super hot and promoted very fast. If the unsorted buffer is full, it is sorted and flushed to \fastdisk{}. There are several levels on \fastdisk{}, following LSM-tree leveling compaction. 
For hotness checking, \ralt{} uses in-memory bloom filters to avoid random disk reads. For calculating range hot set size, \ralt{} stores the prefix sum of the \hotrap{} size of hot records in index blocks.
The details are explained below. For range scans, \ralt{} constructs iterators on each level, like other LSM-trees. To avoid blocking reads, \ralt{} maintains multiple versions of the LSM-tree structure.

If the hot set size or the disk usage of \ralt{} exceeds the limits, an \emph{eviction} is triggered, in which \ralt{} evicts $\beta$ of access records, and all access records are merged into a single sorted run (see below). $\beta=10\%$ is a trade-off between the I/O cost and the stability of \hotrap{}'s hit rate.

\textbf{Calculating scores}.
We use exponential smoothing~\cite{levandoski2013identifying} to calculate scores in order to utilize history information. The score for a key is defined as $\sum_{i=1}^N a_i\alpha^{t-i}$ where $N$ is the number of time slices, $t$ is the current time slice sequence number, and $a_i=1$ if the key is accessed in the $i$-th time slice, $a_i=0$ otherwise. 
In every access record, we store a pair $(tick, score)$ for scoring,
where $tick = t$ is the current time slice sequence number and $score =\sum_{i} a_i\alpha^{tick-i}$. 
\ralt{} maintains the current time slice sequence number $t$, and increments $t$ every time after $\gamma \times (\text{size of \fastdisk{}})$ data have been accessed. 
When the time slice proceeds, we do not need to update the scores in already stored access records; the real score of an access record $(tick, score)$ can be obtained as $\alpha^{t-tick}\cdot score$. 
Two records $(tick_i, score_i), (tick_j, score_j)$ with the same key may need to be merged in \ralt{} which is an LSM-tree. The merged record $(tick^*, score^*)$ becomes $score^*=\alpha^{tick_j-tick_i}\cdot score_i+score_j$ and $tick^*=tick_j$, assuming $tick_i<tick_j$ without loss of generality. We set $\gamma$ to $0.001$ and $\alpha$ to $1-\gamma=0.999$.

\begin{figure}
\centering
\includegraphics{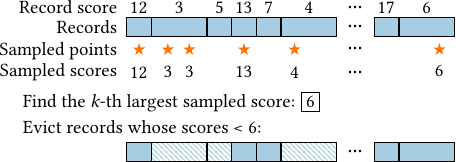}
\vspace{-0.2cm}
\caption{Eviction in \ralt{}. 
\textnormal{The length of records represents their size (\hotrap{} size or physical size). The sampled points are shown as stars in this virtual size space.}
}\label{fig:ralt-eviction-sample}
\end{figure}

\textbf{Eviction}. 
If the hot set size or the physical size of \ralt{} exceeds the limit, \ralt{} evicts $\beta$ records.  
Ideally, we would like to set a score threshold and evict access records with scores smaller than it. However, since the records in \ralt{} are sorted by key, not by score, it is non-trivial to determine a proper threshold. Instead, we use an approximate approach.

We formalize the problem as follows. Each record $i$ has a size $A_i$ (either \hotrap{} size or physical size) and a score $S_i$. We denote the total size as $A = \sum_{i} A_i$. $\beta$ records will be evicted, and the target size after eviction becomes $A'=(1-\beta)A$. 
We want to find a score threshold $S'$, so that records with scores less than $S'$ are evicted, and the total size of records with scores $\ge S'$ is approximately $A'$, i.e., $\sum_{i\in \ralt{}, S_i\geq S'} A_i \approx A'$. 

Our algorithm is illustrated in Figure~\ref{fig:ralt-eviction-sample}.
First, we sample $N$ numbers $a_i$ from the numerical range $[0, A)$, and use a full scan to find the corresponding record $x$ of $a_i$, defined as $\sum_{j<x} A_j\le a_i < \sum_{j\leq x} A_j$; that is, the record that roughly sits at the uniformly sampled position $a_i$ in the space of all records.
Then, the probability for a record $i$ to be sampled is $N A_i / A$.
According to the definition of $S'$, $\sum_{S_i\ge S'} N A_i / A \approx N A' / A$, i.e., the expected number of sampled records with scores $\geq S'$ is $\approx N A' / A = k$.
We thus pick the $k$-th largest score in sampled records to be an approximation of $S'$. 

We use this method to calculate the score thresholds for the two size limits, respectively. 
Records with a score less than the physical size score threshold are evicted from \ralt{}. 
Records with a score less than the hot set score threshold but not less than the physical size score threshold are kept but not considered hot. 
The hotness check and hot key scan operations in \ralt{} ensure this guarantee. 

During eviction, \ralt{} also merges all records. To limit the temporary disk usage, \ralt{} merges step by step. At each step, it picks several unmerged SSTables in the largest level and SSTables that overlap with them in other levels.
It then merges them and updates the version of \ralt{}'s LSM-tree.

\textbf{Checking hotness of a key}.
Searching keys by reading from SSTables is expensive, so for each SSTable, \ralt{} stores bloom filters in memory that contain hot keys (keys with scores higher than the threshold). 
When checking whether a key is hot, \ralt{} checks bloom filters at each level and returns true if any bloom filter gives a positive result. We use 14-bit bloom filters to achieve a low overall false positive rate ($\ll 1\%$). The small number of false positives does not affect performance. Thus we do not perform second-verification.

\textbf{Calculating the hot set size in a range}. 
\ralt{} needs to calculate the hot set size in a given range, for \hotrap{} to select which SSTable to compact (details in \S\ref{subsection:pick-compaction}). Similar to a normal LSM-tree, we have data blocks and index blocks for SSTables. For each data block, its first key and the sum of the \hotrap{} size of hot keys in all previous data blocks are added to the index block. 
For a range query, at each level, we read the two edge index blocks and calculate the difference of their stored sums, to obtain the total \hotrap{} size of hot keys in the range. 
The results of all levels are summed up and returned, as an estimation of the hot set size in the range.

The above result is slightly overestimated because of two error sources.
First, the total \hotrap{} size of hot keys in the one or two edge data blocks is small, so we choose to tolerate the error and not read them. 
Second, there may be duplicate keys in different levels. If the number of levels is small and the size ratio between levels is large, we can expect the overestimation rate to be small. For example, if the size ratio is $10$, the result in the second largest level may be about $1/10$ of the result in the largest level on average. So the overestimation is only about $10\%$.
\S\ref{subsection:pick-compaction} discusses how we handle such overestimation in \hotrap{}.

\textbf{Memory usage and I/O cost}. 
Bloom filters and index blocks are cached in memory. Suppose \hotrap{} record size is 200B and 5\% keys are hot, then the memory usage of bloom filters in \ralt{} is $5\% \times 14/8/200 = 0.0438\%$ of the data size. Suppose the key length is about 20B. Then the size of an index block record is less than 40B. The size of a data block in \ralt{} is 16KiB. Therefore, the memory usage of index blocks is $5\% \times 40/16/1024=0.0122\%$ of the data size. In total, the memory usage is only $0.056\%$ of the data size.

For I/O cost, suppose \ralt{} has $N_L$ levels, and the size ratio is $T$.
Since we evict $\beta$ of data each time, the total write amplification is $\frac{T}{2}N_L+\frac{1}{\beta}$. The total read amplification is $\frac{T}{2}N_L+\frac{2}{\beta}$ because we need two full scans to calculate the score threshold and evict records. In our experiments in \S\ref{sec:cost_breakdown}, $T=10, \beta=10\%, N_L\approx 2$, the read amplification is about $30$ while the write amplification is about $20$. The experiment results show that \ralt{} is only responsible for \RaltIoPortionMin{}--\RaltIoPortionMax{} of total I/O, because a \ralt{} record does not contain the value in a \hotrap{} record, therefore the total I/O of \ralt{} is small compared to \hotrap{}.

\subsection{Checks before inserting to promotion cache}\label{subsection:check-before-promoting-to-cache}

The promotion cache resides between \fastdisk{} and \slowdisk{}. Therefore, before promoting a record into the promotion cache, it is crucial to verify the absence of a newer version of this record in \slowdisk{}, which would otherwise be shielded by the inserted older record in the promotion cache. 
For a point read that wants to insert the latest version just retrieved from \slowdisk{} into the promotion cache, it is known that no newer version exists in the same superversion (i.e., the LSM-tree's snapshot). 
However, it is still possible that a newer version of the record is compacted into \slowdisk{} before the previous record has finished being inserted into the promotion cache.

To address this issue, \hotrap{} marks SSTables as being or having been compacted when setting up compaction jobs.
During the client access, \slowdisk{} SSTables whose range contains the key are recorded. Before inserting a record into the promotion cache,
\hotrap{} checks if any of the recorded SSTables is being or has been compacted. If there is,
\hotrap{} aborts the insertion.
The abort rate is low because of the small number of compaction jobs.
Our experiments show that this check only aborts less than 1\% of insertions into the promotion cache.

\subsection{Concurrency control of promotion by flush}\label{subsection:promotion-concurrent-control}

\begin{figure}
\centering
\includegraphics{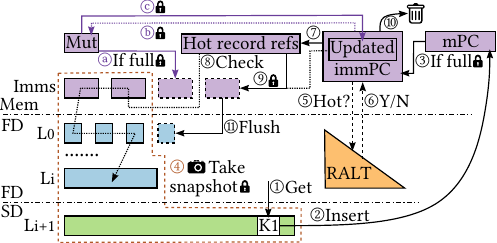}
\vspace{-0.5cm}
\caption{Concurrency control of promotion by flush. \textnormal{Processes with lock icons are protected by the DB mutex lock. \circled{8} ensures that no newer versions exist in the snapshot of the LSM-tree. \circled{a} to \circled{c} insert all updated keys in immutable promotion caches into their \emph{updated} fields. Records with updated keys are excluded in \circled{9}. The snapshot is taken (\circled{4}) after the creation of the immutable promotion cache (\circled{3}), therefore, a key updated before \circled{9} is either found out by \circled{8} or by \circled{a}--\circled{c}.}}\label{fig:promotion-by-flush}
\end{figure}

Now we discuss the details of promotion by flush in Figure~\ref{fig:promotion-by-flush}. 
As mentioned before, a read to \slowdisk{} record (\circled{1}) inserts it to the mutable promotion cache (\circled{2}), which may be later made immutable when full (\circled{3}). 
At this point, a snapshot is taken by incrementing the reference count of the caller's superversion (\circled{4}). We pass the immutable promotion cache's reference and the superversion's reference to a background thread called \emph{Checker}, who handles the rest part. This minimizes the impact on foreground reads.
\emph{Checker} then picks out hot records in the immutable promotion cache by consulting \ralt{} (\circled{5} to \circled{7}). These hot records are candidates to be flushed to $L_0$.

Proper concurrency control is needed to avoid the issue that flushing a hot record to $L_0$ may shield a newer version of this key in the \fastdisk{} levels. 
\emph{Checker} looks for newer versions of the hot records in the snapshot's immutable MemTables and the levels in \fastdisk{} (\circled{8}). 
Hot records without newer versions are packed into a new immutable MemTable (\circled{9})\footnote{Or inserted back into the mutable promotion cache if there are too few of them.}. The immutable promotion cache can be deleted now (\circled{10}). Those records will eventually be flushed into $L_0$ (\circled{11}).

However, there is still a corner case: newer versions can be flushed into $L_0$ in the normal access path when \hotrap{} is looking for newer versions during step \circled{8}. 
To address this issue, \hotrap{} attaches an \emph{updated} field to each immutable promotion cache. When a mutable MemTable becomes immutable during normal accesses (\circled{a}), for every record in it, \hotrap{} checks whether the same key exists in the immutable promotion caches (\circled{b}), and if so, \hotrap{} inserts the key into the \emph{updated} field of the corresponding immutable promotion cache (\circled{c}). The records whose keys exist in the \emph{updated} field will not be packed into the immutable MemTable at step \circled{9}.

To ensure that the list of immutable promotion caches does not change during the creation of immutable MemTables at step \circled{a} so that the \emph{updated} fields are comprehensive, we create immutable promotion caches with the DB mutex (the only major lock in RocksDB) held at step \circled{3}. Since flushes are protected by the DB mutex, there are only two possible cases and both are correct: 
(1) An immutable MemTable is created before an immutable promotion cache is created. The newer versions of records in the immutable MemTable will be detected by the check of \circled{8}. 
(2) An immutable promotion cache is created before an immutable MemTable is created. The newer versions of records in the immutable MemTable will be added to the \emph{updated} field and detected by \circled{a} to \circled{c}.

\subsection{How to pick an SSTable to compact}\label{subsection:pick-compaction}

When a level reaches its capacity, we pick an SSTable and merge it into the next level. Typically, we calculate a score for each SSTable and pick the one with the highest score. For example, RocksDB defines the score as $\emph{FileSize} / \emph{OverlappingBytes}$ by default, in which \emph{OverlappingBytes} represents the total size of the target level's SSTables whose key ranges overlap with the key range of the picked SSTable. This is essentially a cost-benefit trade-off score: \emph{FileSize} is the benefit, and $(\emph{FileSize}+\emph{OverlappingBytes})$ is the cost, in which \emph{FileSize} is optimized out.

However, the cost-benefit score needs some adjustment to better support \hotrap{}. During a cross-tier compaction from \fastdisk{} to \slowdisk{} in \hotrap{}, hot records in the chosen SSTable will be retained in the source level. Therefore, the benefit becomes $(\emph{FileSize} - \emph{HotSize})$, and the cost is still $(\emph{FileSize} + \emph{OverlappingBytes})$. Their ratio is the new cost-benefit score.

\hotrap{} estimates the \emph{HotSize} of an SSTable by querying \ralt{} about the hot set size in the corresponding range (\S\ref{subsection:ralt}).
Recall that the obtained \emph{HotSize} is an overestimation. 
Therefore, it is possible, although very unlikely, that all benefit values are zero. In this case, \hotrap{} falls back to choose the oldest SSTable for compaction.

\subsection{Write amplification of retention}\label{subsection:wa}

Suppose the fraction of cold data in the last level of \fastdisk{} is $p$. 
Since each compaction from \fastdisk{} to \slowdisk{} only compacts $p$ of selected data to \slowdisk{} and writes remaining $(1-p)$ of data back to \fastdisk{}, $\frac{1}{p} \times$ more compactions are needed to compact the same amount of data. Therefore, suppose the size ratio of the LSM-tree is $T$, \fastdisk{} and \slowdisk{} respectively have $n_{\fastdisk{}}$ and $n_{\slowdisk{}}$ levels, the write amplification in \fastdisk{} is $\frac{T}{2}n_{\fastdisk{}} + \frac{1-p}{p}$ and the write amplification in \slowdisk{} is $\frac{T}{2p}+\frac{T}{2}(n_{\slowdisk{}} - 1)$, which are $\frac{1-p}{p}$ and $\frac{T}{2p}-\frac{T}{2}$ larger than a normal LSM-tree, respectively.

Write amplification can be lowered by tuning the level size ratios.
Specifically, we can shrink the first level of \slowdisk{} to make the size ratio between the last level of \fastdisk{} and the first level of \slowdisk{} be $pT$. To keep the size of the LSM-tree in \slowdisk{} unchanged, we can add an extra level after the last level with a size ratio of $\frac{1}{p}$. The size ratio between other levels in \slowdisk{} remains $T$. In this way, the write amplification in \slowdisk{} is $\frac{T}{2} n_{\slowdisk{}} + \frac{1}{2p}$, which is only $\frac{1}{2p}$ larger than a normal LSM-tree.

\subsection{Auto-tuning}\label{subsection:autotune}

In \S\ref{subsection:ralt}, \ralt{} takes 2 parameters: the hot set size limit and the physical size limit, to constrain the total size of hot records and the space of \ralt{} itself, respectively. 
However, the hot set size is determined by the distribution of the workload, which is usually unknown to the user and may further dynamically vary over time.
Hence we propose an automatic tuning approach to estimate the most suitable hot set size and physical size.

We assume each key $k$ has an access probability of $p_k$. For skewed workload distributions, we can set a hot key threshold $p_t$, and find all the keys with $p_k\geq p_t$. When $p_t$ is given, this is a ``frequent items'' problem, with several known solutions like counter-based methods~\cite{LossyCounting2002, SpaceSaving2005} or sketch-based methods~\cite{charikar2002finding}. 
However, counter-based methods need to maintain a large amount of counters, and sketch-based methods also use expensive data structures. 
We proposed a novel algorithm that can dynamically calculate the hot size with small storage overheads and can be implemented by slightly modifying the \ralt{} record format and compaction/eviction strategies.

\begin{algorithm}
	\caption{Algorithm for updating hot set size limit} 
        \label{algo:auto-tune}
        \small
	\begin{algorithmic}[1]
        \State Let the set of tracked keys $T \gets \varnothing$
        \For{each access $k_i$}
            \If{$k_i\in T$}
                \State $c_{k_i}\gets \min(c_{k_i}+\Delta_c,c_{max})$
                \State $t_{k_i}\gets 1$
            \Else{}
                \State $T\gets T\cup \{k_i\}$ 
                \State $c_{k_i}\gets \Delta_c$
                \State $t_{k_i}\gets 0$
            \EndIf
            \If{accessed data amount reaches $R$}
                \State \textbf{forall} $k_i \in T$ \textbf{do} $c_{k_i}\gets \max(c_{k_i}-1, 0)$
            \EndIf
            \If{hot set size limit or physical size limit is exceeded}
                \State Evict $k$ from $T$ with either $c_k$ or $t_k$ is 0
                \State If not enough, evict $k$ with low scores 
                \State {\it\color{gray} \% Update size limits.}
                \State $t\gets \text{\hotrap{} size of stable records}$ \label{algo:auto-tune:update-beg}
                \State $p\gets \text{Physical size of stable records}$
                \State Hot set size limit $\gets \max(L_{hs}, \min(t+D_{hs}, R_{hs}))$
                \State $\text{Physical size limit} \gets p+rD_{hs}$ \label{algo:auto-tune:update-end}
            \EndIf
        \EndFor
	\end{algorithmic} 
\end{algorithm}

Our algorithm to update the two size limits is summarized in Algorithm~\ref{algo:auto-tune}.
We maintain a counter $c$ and a tag $t$ in each \ralt{} access record. Records with $c>0$ and $t=1$ are considered \emph{stable}; otherwise \emph{unstable}. 
The first access to a key sets $c=\Delta_c$ and $t=0$. A hit on an existing key increments $c$ by $\Delta_c$ (capped at $c_{max}$) and activates $t=1$ to be stable. Similar to the score calculation in \S\ref{subsection:ralt}, records with the same keys are merged during compaction. 
Every time when the accessed data amount (in terms of the \hotrap{} size) reaches a limit $R$ (e.g., the size of \fastdisk{}), we decrement all counters by $1$, which eventually makes cold records become no longer stable. 
The maximum counter value $c_{max}$ ensures that cold keys can be evicted after accessing at most $c_{max} \cdot R$ data. 

Old unstable records are evicted when the hot set size or physical size exceeds the limit. 
If the size is still too large, records with low scores will continue to be evicted following the eviction process in \S\ref{subsection:ralt}.
After each eviction, we also update the two limits as depicted in Lines~\ref{algo:auto-tune:update-beg} to \ref{algo:auto-tune:update-end} in Algorithm~\ref{algo:auto-tune}.
Here we have several pre-determined parameters. 
$L_{hs}$ and $R_{hs}$ are the lower and upper bounds of the hot set size; 
$D_{hs}$ is the maximum allowed \hotrap{} size of unstable records. 
Their values in our implementation are listed at the end of this subsection. 
Essentially, we set the hot set size limit to be the size of stable records plus $D_{hs}$, subject to constraints of the two bounds.
Finally, $r$ is the average ratio between physical size and \hotrap{} size, i.e., $\text{\ralt{} access record size} / \text{key-value record size}$.
The physical size limit is thus $r$ times of the hot set size limit.

Our algorithm ensures that if we set $\Delta_c$ and $c_{max}$ carefully, and accesses are identically distributed and independent, then almost all hot keys will become stable, while the size of stable cold keys is bounded. 
We have formally proved these properties, but we omit the proof due to space limit.
Intuitively, because of the tag $t$, we need to access an originally unstable key at least two times with less than $D_{hs}$ other data accessed in between, in order to make it stable during $R$ amount of data accessed. 
While the difference between $\sum_{k \text{ is hot}} p_k$ and $\sum_{k \text{ is cold}} p_k$ may not be large enough due to the very large number of cold keys, the difference between $\sum_{k \text{ is hot}} p_k^2$ and $\sum_{k \text{ is cold}} p_k^2$ is typically large enough to separate them apart. 

\textbf{Implementation details}.
We set $L_{hs}=0.05 \times \text{\fastdisk{} size}$, $R_{hs}=0.7 \times \text{\fastdisk{} size}$, $D_{hs} = 0.1 \times R_{hs}$, $R=R_{hs}$. We set $\Delta_c=2.6, c_{max}=5$, $p_t=\frac{1}{0.5 \times \text{\fastdisk{} size}}$. 
With these parameters, according to our analysis, the \hotrap{} size of cold keys with $p_k < 0.1 \times p_t$ is smaller than $0.5 \times \text{\fastdisk{} size}$, which is acceptable. The probability of each hot key becoming stable is at least $98\%$. The experiments are shown in \S\ref{evaluation:dynamic-workload}.

\textbf{Limitation}.
The algorithm keeps unstable records of \hotrap{} size $D_{hs}$ to detect hot records. It assumes that hot records are accessed randomly, so that every hot record has a probability $>0$ to be detected. If a hot record is accessed once every $>D_{hs}$ accesses, then it cannot be captured. In the extreme case, such as ``sequential flooding'', the algorithm cannot work unless increasing $D_{hs}$. But this is uncommon in real-world datasets (for example, the Twitter trace in \S\ref{evaluation:twitter}).

\section{Evaluation}\label{section:eval}

\subsection{Experimental setup}

\begin{table}
\centering
\caption{Disk performance on our EC2 instances.}\label{tab:aws-configs}
\vspace{-0.3cm}
\tablefontsize
\begin{tabular}{|c|c|c|}
\hline
& \fastdiskfull{} & \slowdiskfull{} \\
\hline
Type & AWS Nitro SSD & gp3 \\
\hline
16 threads rand 16K read IOPS & $\approx$83000 & 10000 \\
\hline
Sequential read bandwidth & $\approx$1.4GiB/s & 1000MiB/s \\
\hline
Sequential write bandwidth & $\approx$1.1GiB/s & 1000MiB/s \\
\hline
\end{tabular}
\end{table}

\begin{table}
\caption{Read-write ratios of YCSB workloads in our tests.}\label{tab:read-write-ratios}
\vspace{-0.3cm}
\tablefontsize
\centering
\begin{tabular}{|c|c|c|}
\hline
Notation & Meaning & Read-write ratio \\
\hline
RO & read-only & 100\% read \\
\hline
RW & read-write & 75\% read, 25\% insert \\
\hline
WH & write-heavy & 50\% read, 50\% insert \\
\hline
UH & update-heavy & 50\% read, 50\% update \\
\hline
\end{tabular}
\end{table}

\begin{figure*}[t!]
	\centering
	\includegraphics{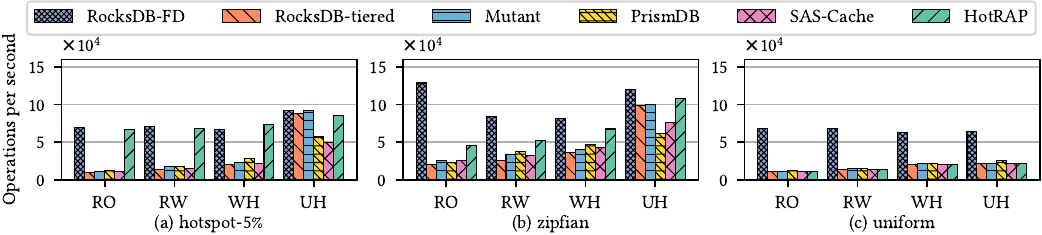}
    \vspace{-0.3cm}
	\caption{Throughput comparison with 1KiB record size.}\label{fig:ycsb-sweep}
\end{figure*}

\noindent\textbf{Testbed}. We evaluate HotRAP on AWS EC2 i4i.2xlarge instances running Debian 12. Each instance has 8 vCPU cores, 64GiB memory, and a 1875GB local AWS Nitro SSD. 
We use local SSDs as \fastdisk{} and gp3 as \slowdisk{}, with their performance shown in Table~\ref{tab:aws-configs}. 
Since HDD RAID arrays can achieve thousands of IOPS and high throughput~\cite{WikiRAID}, we set the maximum IOPS and throughput of gp3 to 10000 and 1000MiB/s respectively to simulate the most performant HDD RAID arrays.

\noindent\textbf{Sizes of tiers}. We set the space ratio between tiers to 10: the expected used size of \fastdisk{} is set to 10GB and the initial expected used size of \slowdisk{} is set to 100GB.

\noindent\textbf{Compared systems}. We compare \hotrap{} with Mutant~\cite{yoonMutantBalancingStorage2018}, PrismDB~\cite{PrismDB}, SAS-Cache~\cite{SAS-Cache}, and following two variants of RocksDB. \emph{RocksDB-FD} stores all data in \fastdisk{}, which is used to indicate the upper-bound performance that \hotrap{} can achieve.
\emph{RocksDB-tiered} tunes its size ratios between levels so that the total size of \fastdisk{} levels becomes 10GB, which is the same as \hotrap{}. Mutant, PrismDB, and SAS-Cache are also tuned to use about 10GB of \fastdisk{} expectedly.

\noindent\textbf{Configurations}. All experiments run with 16 threads.
The block size of all systems is set to 16KiB following Meta's practice~\cite{RocsDBWikiMem}. The maximum number of background jobs is set to 6 for \hotrap{} and RocksDB. To minimize the impact of the file system cache, direct I/O is used for \hotrap{}, RocksDB, and PrismDB, for fair comparison. For Mutant, since it is difficult to enable direct I/O due to its old version of RocksDB, we limit its memory with \verb|systemd-run| to reduce the page cache size. Auto-tuning of HotRAP is turned on. We set the initial hot set size limit and \ralt{} physical size limit to 50\% and 15\% of the \fastdisk{} size respectively. The secondary cache size of SAS-Cache is set to 6GB. \hotrap{} is configured with a 128MiB block cache, while other systems are configured with 64MiB more block cache to compensate for the memory usage of \ralt{}. All systems are configured with 10-bit bloom filters. Other configurations are set to their default values.

\subsection{Performance on YCSB workloads}

\begin{figure}[t!]
	\centering
	\includegraphics{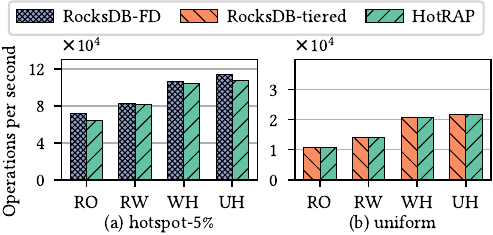}
    \vspace{-0.3cm}
	\caption{Throughput comparison with 200B record size.}\label{fig:ops-200B}
\end{figure}

\begin{figure}[t!]
	\centering
	\includegraphics{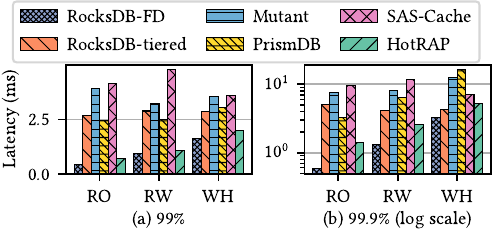}
    \vspace{-0.3cm}
	\caption{\emph{Get} tail latency comparison under hotspot-5\% workloads with 1KiB record size.}\label{fig:latency}
\end{figure}

To assess the performance of \hotrap{} under various key distributions and read-write ratios, we evaluate \hotrap{} with the YCSB workloads~\cite{ycsb} shown in Table~\ref{tab:read-write-ratios}. \emph{RO} (read-only) tests the effectiveness of promotion.
\emph{RW} (read-write) and \emph{WH} (write-heavy) test the effectiveness of retention. \emph{UH} (update-heavy) is the worst case for \hotrap{}. With UH, the key distributions of reads and updates are the same. Therefore, newer versions of read-intensive records are frequently inserted into the database and flushed into \fastdisk{}, making the proactive promotion of \hotrap{} barely needed.

We test three skewness types: hotspot-5\%, Zipfian, and uniform.
In the hotspot-5\% distribution, 5\% of records are accessed by 95\% operations uniformly. The other 5\% operations uniformly access the other 95\% of records.
In the Zipfian distribution, the access probability of the $k$-th hottest record is proportional to $1 / k^s$~\cite{ZipfWiki}. In our experiments, $s = 0.99$. In the uniform distribution, the access probability of all records is the same.
We evaluate two record sizes, 1KiB ($\approx$24B key and 1000B value) and 200B ($\approx$24B key and 176B value).

All workloads consist of the load phase and the run phase. The load phase loads 110GB of records into the LSM-tree. The run phase executes read/write operations. We focus on the performance of the run phase.
For workloads with 1KiB records, the run phase executes $2.2\times 10^8$ operations. For 200B records, the run phase executes $1.1\times 10^9$ operations.

Figures~\ref{fig:ycsb-sweep} and \ref{fig:ops-200B} compare the average throughput (over the final 10\% of the run phase) of evaluated systems under different read-write ratios and skewness types, with 1KiB and 200B records, respectively. 
Since the trends are similar, we only show a representative subset in Figure~\ref{fig:ops-200B} to save space.

Under both record sizes, the performance of \hotrap{} when running hotspot-5\% workloads is close to that of the ideal RocksDB-FD, because \hotrap{} promotes almost all hot data into \fastdisk{} and achieves about $95\%$ hit rate.
Compared to other systems, \hotrap{} achieves \SpeedupRO{}$\times$ speedup over the second best for read-only (RO) workloads, and \SpeedupRW{}$\times$ for read-write-balanced workloads (RW).
On the other hand, \hotrap{} is only \OverheadUniformRocksdbTiered{} slower than RocksDB-tiered under uniform workloads, showing the overhead of \hotrap{} is low when promotion has no benefits.
For the Zipfian distribution, there is a noticeable gap in Figure~\ref{fig:ycsb-sweep} between \hotrap{} and the upper-bound RocksDB-FD, because the hit rate is lower ($79\%$) compared to hotspot-5\%. But \hotrap{} still outperforms other designs. 

For the three baseline designs, Mutant and SAS-Cache show negligible improvements compared to RocksDB-tiered, because their data promotion granularity is too coarse: data blocks for SAS-Cache and SSTables for Mutant. PrismDB also only has small improvements over RocksDB-tiered due to its inefficient promotion mechanism. 

Among different read-write ratios, we notice that all systems exhibit higher throughput under the update-heavy (UH) workloads. This is because the updated data have a skewed distribution, and thus update operations can be considered as promotions to \fastdisk{}. 
However, PrismDB and SAS-Cache perform relatively poorly in this case. PrismDB suffers from lock contention and inefficient random writes to \fastdisk{}. SAS-Cache has its secondary cache frequently evict and promote data blocks, thus incurring significant \fastdisk{} I/O.

Figure~\ref{fig:latency} shows the tail latencies (again over the final 10\% of the run phase) under hotspot-5\% workloads with 1KiB record size.
When running read-heavy workloads (RO \& RW), \hotrap{} achieves lower tail latency than other systems except RocksDB-FD. 
In these cases, \hotrap{} has higher \fastdisk{} hit rates and thus reduces accesses to \slowdisk{}, so there are proportionally fewer long-latency accesses affected by the \slowdisk{} tail latency among all \fastdisk{} and \slowdisk{} accesses. 

\subsection{Performance on real-world Twitter traces}\label{evaluation:twitter}

\begin{figure}[t!]
	\includegraphics{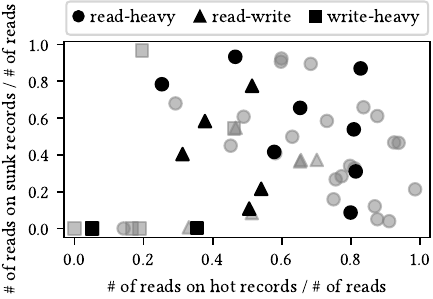}
    \vspace{-0.3cm}
	\caption{Characteristics of Twitter production traces. \textnormal{Each point stands for a cluster's trace. Dark black points are our selected traces for evaluation in Figure~\ref{fig:twitter-speedup}.}}\label{fig:twitter-scatter}

    \vspace{0.4cm}

    \includegraphics{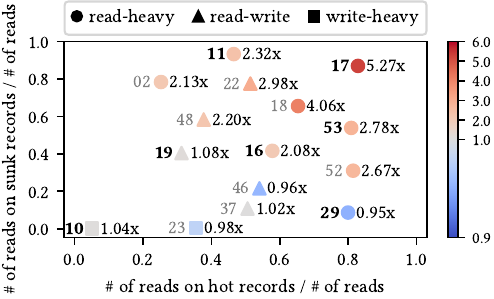}
    \vspace{-0.3cm}
	\caption{Speedup of \hotrap{} over RocksDB-tiered on Twitter production traces. \textnormal{Numbers on the two sides of a point are the cluster ID and the speedup. Traces with bold cluster IDs are selected for further analysis in Figure~\ref{fig:twitter-ops}.}}\label{fig:twitter-speedup}
\end{figure}

\begin{figure}[t!]
	\centering
	\includegraphics{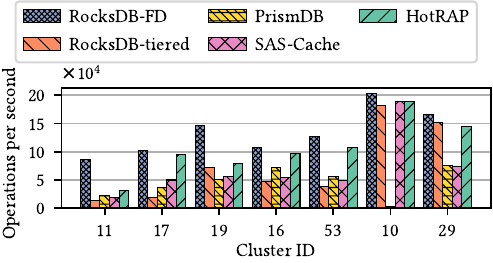}
    \vspace{-0.3cm}
	\caption{Throughput comparison under several Twitter production traces.}\label{fig:twitter-ops}
\end{figure}

To show the performance of \hotrap{} under real-world workloads, we evaluate \hotrap{} with the Twitter production traces~\cite{twittertrace}. We pre-process every trace into two phases: the load phase and the run phase. In the load phase, we ignore reads and keep inserting about 110GB of data. In the run phase, we execute $5\times 10^8$ operations. We augment small traces by repeating operations until reaching 110GB of data. For example, if a trace has 40GB of data, we repeat each operation 3 times, e.g., an operation ``\verb|INSERT user1|'' becomes ``\verb|INSERT 0user1, INSERT 1user1, INSERT 2user1|''.

Traces are categorized based on their read proportions: read-heavy has $>$75\% reads; read-write has $>$50\% and $\le$75\% reads; write-heavy has $\le$50\% reads. 
In some traces, it is common for a frequently read key to be also frequently updated. Proactively promoting this key has little benefit because its newer version will be automatically inserted into \fastdisk{}. 
We define that a read is performed on a \emph{sunk} record if the data amount written since the last update of its key is greater than 5\% of the DB size, so that the latest version is now likely sunk to \slowdisk{} when being read.
For example, suppose we first update keys $A, C, D, B$, then we read key $A$. If the total size of records $C, D, B$ is greater than 5\% of the DB size, then we say $A$ is sunk when being read.
Similarly, we define that a read is performed on a \emph{hot} record if the data amount read since the last read of this record is less than 5\% of the DB size, so that the record is likely to be identified as hot. 
If there are many reads on such sunk and hot records, then it is necessary to promote them to \fastdisk{}.

Figure~\ref{fig:twitter-scatter} shows the categories and proportion of reads on sunk/hot records in the Twitter traces. 
We select some representative traces to evaluate the speedup of \hotrap{} over RocksDB-tiered in Figure~\ref{fig:twitter-speedup}. 
\hotrap{} performs better when the proportions of reads on sunk and hot records are higher, e.g., achieving up to \MaxSpeedupTwitterRocksdbTiered{}$\times$ speedup under the trace of cluster 17. 
On the other hand, \hotrap{} is not significantly slower than RocksDB-tiered under traces with low proportions of reads on sunk and hot records, showing its low overhead.

We further analyze several traces with high (11, 17), middle (19, 16, 53), and low (10, 29) proportions of sunk record reads and show their throughput in Figure~\ref{fig:twitter-ops}. 
Mutant is not shown because it incurs too much FD usage ($>$30GB) when executing several traces.
\hotrap{} is almost always the best among compared systems, and achieves up to \MaxSpeedupTwitter{}$\times$ speedup over the second best.
When the proportion of sunk record reads is lower, almost all systems perform better, but the speedup of \hotrap{} also becomes smaller. At middle and high proportions of sunk record reads, if we increase the amount of hot record reads, \hotrap{} throughput would significantly increase, while PrismDB and SAS-Cache perform moderately better, and RocksDB-tiered does not benefit from it at all.

PrismDB performs extremely badly under cluster 10 trace because this trace uniformly reads recently updated keys which are still in \fastdisk{}. Therefore PrismDB identifies almost all data in \fastdisk{} as hot and retains them. As a result, when \fastdisk{} gets full, PrismDB can only demote a few keys to \slowdisk{}, and it has to trigger demotion repeatedly to keep \fastdisk{} usage under the limit. The frequent demotion contends for locks with reads and thus causes severe performance degradation.

\subsection{Cost breakdown}\label{sec:cost_breakdown}

\begin{figure}[t!]
	\centering
	\includegraphics{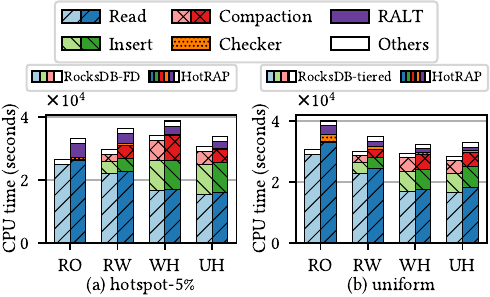}
    \vspace{-0.3cm}
	\caption{CPU time breakdown with 200B record size.}\label{fig:breakdown-cpu-time-200B}
\end{figure}

\begin{figure}[t!]
	\centering
	\includegraphics{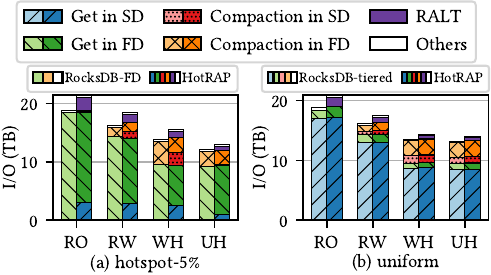}
    \vspace{-0.3cm}
	\caption{I/O breakdown with 200B record size.}\label{fig:breakdown-io-200B}
\end{figure}

\begin{figure*}[t!]
\includegraphics{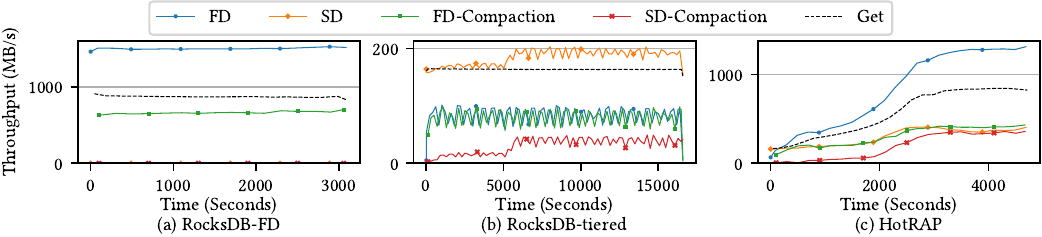}
\vspace{-0.3cm}
\caption{Device throughput breakdown under the RW (75\% read, 25\% insert) hotspot-5\% workload with 1KiB record size. 
\textnormal{For readability, throughput is averaged every 200 seconds. 
\emph{Get} is the measured throughput of disk reads during \emph{Get} operations, which is an indicator of end-to-end performance.}}\label{fig:breakdown-device-throughput}
\end{figure*}

Figures~\ref{fig:breakdown-cpu-time-200B} and \ref{fig:breakdown-io-200B} show the breakdown of CPU time and I/O for the run phase of YCSB workloads with 200B record size.
The size of \ralt{} here is more than 1GB, which needs to be stored in \fastdisk{} instead of memory.
The results show that \ralt{} is only responsible for \RaltCpuPortionMin{}--\RaltCpuPortionMax{} of total CPU time and \RaltIoPortionMin{}--\RaltIoPortionMax{} of total I/O, hence the design is efficient.

In the hotspot-5\% workloads, \hotrap{} incurs more CPU time and I/O on compaction than RocksDB-FD, because retention increases write amplification.
In the uniform workloads, \hotrap{} consumes more CPU time than RocksDB-tiered because most accesses are in \slowdisk{} and thus the records are inserted into the promotion cache. However, few records are promoted into \fastdisk{} due to the hotness checking (\circled{7} \& \circled{b} in Figure~\ref{fig:overview}), therefore they have similar compaction I/O.

Figure~\ref{fig:breakdown-device-throughput} shows the breakdown of device throughput under the RW (75\% read, 25\% insert) hotspot-5\% workload. 
\hotrap{} promotes and retains hot records in \fastdisk{}. Therefore, the number of \emph{Get} operations served by \fastdisk{} increases over time, raising the total throughput of \emph{Get} until it is close to that of RocksDB-FD. 
Most \emph{Get} operations in RocksDB-tiered are served by \slowdisk{}, which bounds its total \emph{Get} throughput.

\subsection{Effectiveness of individual techniques}\label{evaluation:eff}

\begin{table}[t!]
	\centering
	\caption{Promotion costs with/without retention under the RW hotspot-5\% workload with 1KiB record size.}\label{tab:no-retain}
    \vspace{-0.3cm}
    \tablefontsize
	\begin{tabular}{|c|c|c|c|c|}
	\hline
	Version & Promoted & Retained & Compaction & Hit rate \\
	\hline
	HotRAP & 6.2GB & 40.2GB & 2346.1GB & 94.5\% \\
	\hline
	no-retain & 35.1GB & 0.0GB & 3038.6GB & 71.4\% \\
	\hline
\end{tabular}

\end{table}

To show the effectiveness of retention, we remove the retention mechanism from \hotrap{} and call the design \emph{no-retain}. Table~\ref{tab:no-retain} shows that \emph{no-retain} incurs more promotion costs and achieves a lower final hit rate than \hotrap{}. The reason is that although \emph{no-retain} still promotes records into \fastdisk{}, the promoted records are compacted into \slowdisk{} again during compactions. Therefore, hot records have to be promoted repeatedly, with much more promotion traffic.

\begin{table}[t!]
	\centering
	\caption{Promotion costs with/without hotness checking under the RO uniform workload with 1KiB record size.}\label{tab:promote-accessed}
    \vspace{-0.3cm}
    \tablefontsize
	\begin{tabular}{|c|c|c|c|}
	\hline
	Version & Promoted & Retained & Compaction \\
	\hline
	HotRAP & 1.2GB & 0.0MB & 26.6GB \\
	\hline
	no-hotness-check & 205.8GB & 9.0GB & 6645.1GB \\
	\hline
\end{tabular}

\end{table}

\hotrap{} only promotes hot records to reduce the overhead introduced by promotion. To show its effectiveness, we remove hotness checking and promote all accessed records. Table~\ref{tab:promote-accessed} shows that under uniform workloads, \emph{no-hotness-check} promotes \PromoteAccessedPromotedMore{} more records and thus incurs \PromoteAccessedCompactionMore{} more compaction I/O than \hotrap{}.

\subsection{Performance on dynamic workload}\label{evaluation:dynamic-workload}

\begin{figure}[t!]
	\centering
	\includegraphics{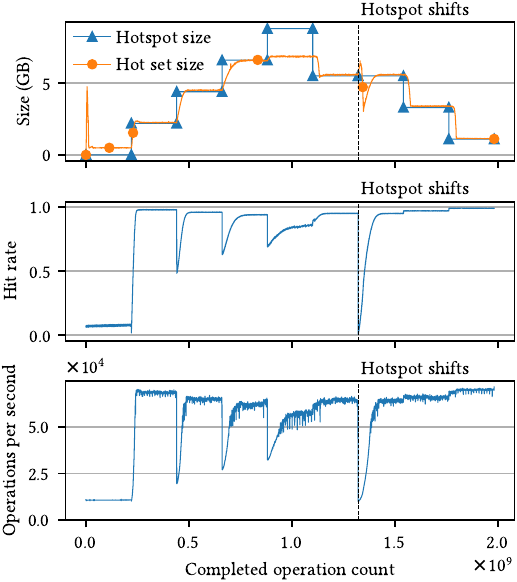}
    \vspace{-0.7cm}
	\caption{\hotrap{} under dynamic workload. \textnormal{The run phase consists of nine stages, whose data distributions are first uniform, then hotspot-2\% $\to$ 4\% $\to$ 6\% $\to$ 8\% $\to$ 5\% $\to$ 5\% $\to$ 3\% $\to$ 1\%. Each stage executes $2.2\times 10^8$ read operations.
    The two 5\% hotspots in the 6th and 7th stages are non-overlapping. 
    When the hotspots increase from 2\% to 8\%, the new hotspot completely contains the old hotspot. When they decrease, the new one is completely contained by the old one.}}\label{fig:dynamic-workload}
\end{figure}

To show that \hotrap{} can adapt to changes in the access pattern, we evaluate it under a dynamic workload. The details of the dynamic workload and the results are shown in Figure~\ref{fig:dynamic-workload}. 

The hot set starts from 5GB.
The first stage has a uniform distribution, so there are few stable records, and the hot set size limit decreases to the minimum value 0.5GB.
At the second stage, the key distribution becomes hotspot-2\%. With the auto-tuning method in \S\ref{subsection:autotune}, \hotrap{} gradually increases the hot set size limit until all hotspot keys are added to the hot set. Eventually, the hot set size stabilizes around the hotspot size.
After the hotspot expands from 2\% to 4\% and from 4\% to 6\%,
the \fastdisk{} hit rate temporarily drops because new hot keys are not yet promoted. Then \hotrap{} gradually increases the hot set size limit until all new hot keys are added to the hot set, thus recovering the hit rate.
After the hotspot expands from 6\% to 8\%, the hotspot size exceeds the max hot set size (7GB). Therefore, the performance of \hotrap{} is relatively low.
When the hotspot shifts, \hotrap{} reacts adaptively by evicting old hot keys after they become unstable, and gradually adding new hot keys to the hot set. Both the throughput and the hit rate recover eventually.
After the hotspot shrinks from 5\% to 3\% and from 3\% to 1\%, the throughput and the hit rate do not drop because new hot keys are already considered hot and have already been promoted. Nevertheless, \hotrap{} in this case could decrease the hot set size limit after old access records become unstable.

In summary, the results show that the auto-tuning mechanism enables \hotrap{} to find the most suitable hot set size limit under a dynamic workload with hotspot expanding, shrinking, and shifting.

\section{Conclusion}
We introduced \hotrap{}, an LSM-tree-based key-value store on tiered storage. Unlike previous solutions, \hotrap{} adopts an efficient on-disk hotness tracker, along with fine-grained record-level retention and promotion mechanisms that independently operate besides LSM-tree compactions. These techniques allow \hotrap{} to move data efficiently across tiers to fully utilize the fast storage for keeping hot records even under read-heavy workloads.

\bibliographystyle{plain}
\bibliography{hotrap}

\end{document}